\documentclass[fleqn,twoside]{article}%
\usepackage{amsmath}
\usepackage{amsfonts}
\usepackage{amssymb}
\usepackage{graphicx}%
\def\be{\begin{equation}}

\def\ee{\end{equation}}
\def\bes{\begin{equation}\begin{split}&}
\def\es{\end{split}}
\def\bi{\bibitem}

\begin{document}
\title{Semiclassical gravity with a nonminimally coupled scalar field.}
\author{Abhik Kumar Sanyal}

\maketitle
\noindent
\begin{center}
\noindent
Dept. of Physics, Jangipur College, Murshidabad, West Bengal, India-742213\\
and\\
Relativity and Cosmology Research Centre, Dept. of Physics, Jadavpur University, Calcutta, India-700032\\
\end{center}
\footnotetext[1]
{\noindent
Electronic address:\\
\noindent
sanyal\_ak@yahoo.com}

\begin{abstract}
Semiclassical approximation to the Wheeler-DeWitt equation which corresponds to gravity with a minimally coupled scalar field has been performed. To the leading order, vacuum Einstein's equation along with the functional Schr\"odinger equation for the matter field, propagating in the background of classical curved space are obtained. The Schr\"odinger equation is solved for a quartic potential. It is observed that the wave-functional admits the wormhole boundary condition even for large negative values of the coupling constant $\epsilon$. For conformal coupling $\epsilon = {1\over 6}$, the Hawking-Page wormhole solution is recovered. \\
\end{abstract}

\section{Introduction}
It has been well established \cite{1, 2, 3, 4} that much below Planck's scale, the Wheeler-DeWitt equation for gravity in the presence of a minimally coupled scalar field can be approximated to obtain equations corresponding to quantum field theory in curved space-time, along with the vacuum Einstein's equation with possible back reaction. For this purpose, one has to expand the WKB phase of the wave-functional $\psi$ in the power series of Planck's mass $m_{pl}$, rather than the Planck's constant $\hbar$. This method of approximation has also been successfully applied by the present author \cite{5} to a recent theory - ``A Non-singular Universe" \cite{6, 7} - which is essentially a constrained system of the non-standard cosmological model that includes a higher order curvature invariant term. So far there has been little effort to apply the said method of approximation to gravity with non-minimally coupled scalar field.\\

In recent years it has been observed that the non-minimal coupling might play an important role in improving the inflationary models \cite{8, 9, 10, 11}. On one hand the extended inflation obtained by La and Steinhardt \cite{12} based on the solutions obtained by Mathiazhagan and Johri \cite{13} in the Brans-Dicke theory of gravity and that obtained by Accetta and Trester \cite{14} in induced theory of gravity, raised the hope that non-minimally coupled scalar field might help to solve the ``graceful exit" problem suffered by most of the inflationary models. On the other hand, recent works of Fakir and Unruh \cite{15} and also of Fakir, Habib and Unruh \cite{16} suggest that non-minimal coupling might help solving the long standing problem of density perturbation.\\

One of the greatest successes of the inflationary models is that these can give rise to initial perturbations which can provide explanation for structure (galaxy, clusters, super-clusters etc.) formation. However, the trouble with all the inflationary models is that, they give rise to density perturbations which are five to six order of magnitudes in excess of that required. In order to fit the observed density perturbation, the self consistent parameter $\lambda$ of the scalar field is usually tuned to an extremely low value ($\lambda \le 10^{-12}$) which is about ten order of magnitude smaller than the values calculated from standard field theory.\\

In a couple of papers \cite{15, 16} Fakir, Unruh and Habib have shown that instead of tuning the self coupling parameter $\lambda$, it is indeed possible to solve this long standing problem in principle, simply by constraining the non-minimal coupling constant $\epsilon$ to a large negative value ($\sim -10^{-3}$). The outcome of their analysis has proved that the large negative $\epsilon$ may lead to well-behaved, self-consistent classical solutions which admit inflation, despite previous belief to the contrary. Further, they have demonstrated rigorously that within the frame work of chaotic inflation \cite{11}, one can produce density perturbations of amplitudes consistent with the large scale behaviour, keeping $\lambda$ within the order ($\sim 10^{-2}$) of the ordinary GUT range.\\

These encouraging aspects of non-minimally coupled scalar fields led Fakir \cite{17} to study it in the context of quantum cosmology. He obtained the first order WKB solutions to the modified Wheeler-DeWitt equation and showed that the generic features of Vilenkin \cite{18} Hartle-Hawking \cite{19} wave-functions remain preserved in non-minimally coupled case. He also tried to present a probabilistic interpretation of the wave-function by arguing that instead of being the wave-function of the universe, the minisuperspace wave-function could describe the quantum creation of inflationary sub-universes in Linde's chaotic inflationary model \cite{11}.\\

So far we have discussed the success of the non-minimal coupling in the context of the inflationary scenario. However, interest in the field has increased following a recent paper of Coule \cite{20}, who has shown that a non-minimally coupled scalar field admits wormhole solutions both for real and imaginary fields.\\

Wormholes \cite{21, 22, 23, 24} are usually treated as solutions to the classical Euclidean field equations. For such wormholes to exists, the Ricci tensor ($R_{\mu\nu}$) should have negative eigenvalues, and as such classical wormholes do not exist for minimally coupled real massless scalar fields. The situation changes in the case of non-minimal coupling ($\epsilon \ne 0$), since in that case $R_{\mu\nu} \simeq (M_{pl} - 6\epsilon \phi^2) \partial_{\mu}\phi \partial_{\nu}\phi$, where, $M_{pl}$ is the Planck's mass. Therefore, classical wormholes can exist only for positive values of $\epsilon$, provided the field takes some appropriate large value to make $R_{\mu\nu} < 0$. Hawking and Page \cite{25} have shown that wormholes can be represented in a more general manner as solutions of Wheeler-DeWitt equations with appropriate boundary conditions. Such boundary conditions can be formulated sketchily in the following manner. The wave-function should be damped for large three-volume, while it should tend to a constant value for small three-volume and of-course should be regular everywhere. Hawking and Page \cite{25} have also produced such quantum wormhole solutions for massless minimal and conformal scalar fields. Coule \cite{20} on the other hand, studied wormhole solutions for non-minimally coupled scalar fields from the view point of the sign of the potential term appearing in the Wheeler-DeWitt equation, instead of solving the equations explicitly. The outcome of his work is that, like classical wormholes, quantum wormholes also exist, but for positive value of $\epsilon$ in the case of real massless scalar field. However, for an imaginary scalar fields, wormholes exist for all positive and negative values of $\epsilon$, and of course for $\epsilon = 0$, i.e. for minimally coupled fields.\\

Since macroscopic wormholes are supposed to provide a mechanism for the evaporation and complete disappearance of black-holes, while microscopic wormholes might play an important role for the vanishing of the cosmological constant, it is potentially an important scenario rather than a model, that might have played a dominant role in the evolution of the very early universe. Nevertheless, the fact that only positive value of $\epsilon$ seems to provide wormhole solutions for a real massless scalar field, while large negative $\epsilon$ can solve the density perturbation problem, is surely worrisome. However, the issue of wormholes has not been settled as yet. Coule \cite{20} has suggested that while relating quantum wormholes to the classical ones, it might predict a minimum size of the proper volume of the universe, below which the wave-function would vanish.\\

In view of the above discussions, we are motivated to study the cosmological aspect of the theory of gravity when it is non-minimally coupled to a real scalar field, yet from a new direction. As already mentioned, here we apply semi-classical technique to the corresponding Wheeler-DeWitt equation by expanding the phase of the wave-function in the power series of the Planck's mass. In the process we have derived the Tomonaga-Schwinger equation, which is essentially the functional Schr\"odinger equation for the scalar field propagating in the background of classical curved space. We then solve it under the choice of quartic potential $V(\phi) = \lambda \phi^4$, and ultimately presented the wave-functional with higher order correction term. Next, we are motivated to check whether our solution satisfies the wormhole boundary condition for large negative value of $\epsilon \sim (-10^{-3})$. We have in the process confirmed that at least one such wormhole solution exist for a real free scalar field, and one for a self-coupled field in the closed ($k = + 1$) universe model.

\section{Semiclassical Wave-functional and Wormhole Configuration}

We start with the following non-minimally coupled gravitational action,

\be\label{2.1} S = \int d^4 x \sqrt{-g} \left[-{1\over 16\pi G}\left(1 - {4G\over \pi}\epsilon \phi^2\right) R - {1\over 2\pi^2}\left({1\over 2}\phi_{,\mu}\phi^{\mu}+ V(\phi)\right)\right] - {1\over 8\pi G}\int_{\Sigma}d^3 x\sqrt{h} K\left(1 - {4G\over \pi}\phi^2\right),\ee
where, $\epsilon$ is the coupling constant, and while $\epsilon = 0$ corresponds to minimal coupling, $\epsilon = {1\over 6}$ corresponds to conformal coupling. The surface term contains the determinant of the three-metric $h$ and the trace $K$ of the extrinsic curvature tensor $K_{ij}$. In the Robertson-Walker line element

\be\label{2.2} ds^2 = -dt^2 + a(t)^2\left[{dr^2\over 1-kr^2} + r^2(d\theta^2 + \sin^2{\theta}~d\phi^2)\right]\ee
the action \eqref{2.1} reduces to

\be\label{2.3} S = \int\left[(M - 6\epsilon\phi^2)\left(-{1\over 2}a\dot a^2 +{k a\over 2}\right)+ 6\epsilon a^2\phi~\dot a~\dot\phi + \left({1\over 2}\dot\phi^2 - V(\phi)\right)a^3\right]dt,\ee
where, $M = {3\pi\over 2G} = {3\over 2}\pi m_{pl}^2$, $m_{pl}$ being the Planck's mass. Thus the $\left(^0_0\right)$ component of Einstein's field equations, viz. the classical Hamiltonian constraint equation in configuration space variables reads

\be\label{2.4} -{1\over 2}\left({\dot a^2\over a^2} + {k\over a^2}\right) + {1\over M - 6\epsilon\phi^2}\left[6\epsilon{\dot a\over a}\phi\dot\phi+{1\over 2}\dot\phi^2 + V(\phi)\right]=0.\ee
In terms of phase-space variables, the corresponding Hamiltonian constraint equation reads

\be\label{2.5} {1\over M - 36\epsilon^2\phi^2- 6\epsilon\phi^2}\left(-{p_a^2 \over 2a} + {M - 6\epsilon\phi^2\over 2 a^3}p_{\phi}^2 + {6\epsilon\phi\over a^2}p_ap_{\phi}\right) - {M - 6\epsilon\phi^2\over 2}ka + a^3 V(\phi) = 0.\ee
To follow the standard canonical quantization scheme, we can now express the Wheeler-DeWitt equation \eqref{2.5} after carefully removing the operator ordering ambiguities between $\hat a$ and $\hat{p_a}$, $\hat\phi$ and $\hat{p_{\phi}}$. For this purpose, we replace $\hat p_a$ by $a^{-n}\hat p_a a^n$, $\hat p_a^2$ by $a^{-n}\hat p_a a^n \hat p_a$; $\hat p_{\phi}$ by $\phi^{-l}\hat p_{\phi}\phi^l$, and $\hat p_{\phi}^2$ by $\phi^{-l}\hat p_{\phi}\phi^l\hat p_{\phi}$, where $n$ and $l$ are operator ordering indices. Next, momentum operators are replaced by corresponding gradient operators and thus the Wheeler-DeWitt equation reads

\be\begin{split}\label{2.6}
&\Bigg[ \hbar^2 \left\{ {1\over 2a}{\partial^2\over \partial a^2} - {M-6\epsilon\phi^2\over 2 a^3}{\partial^2\over \partial\phi^2}+{1\over a^2}\left({n\over 2}-6\epsilon l\right){\partial \over \partial a} - {1\over 2a^3\phi}\big[12 \epsilon n \phi^2 - (M - 6\epsilon\phi^2)l\big]{\partial\over \partial \phi}-{6\epsilon\phi\over a^2}{\partial^2\over \partial a\partial\phi} - {6\epsilon n l\over a^3}\right\}\\
& - (M +36\epsilon^2\phi^2 - 6\epsilon\phi^2)\left({M-6\epsilon\phi^2\over 2}k a - a^2 V(\phi)\right)\Bigg]|\psi> = 0
\end{split}\ee
We can now carry out semi-classical approximation by writing $\psi$ as $\psi = \exp\left({iS\over \hbar}\right)$ and expanding the phase in the power series of $M$ as $S = MS_0 + S_1 + M^{-1} S_2 + ....$ etc. Finally, substituting all these in the above Wheeler-DeWitt equation \eqref{2.6}, and comparing expressions having the same order in $M$, we obtain to the highest order $M^3$, the following equation,

\be\label{2.7} {\partial S_0\over \partial \phi} = 0,\ee
which clearly states that $S_0$ is purely a function of the gravitational field variable. To the next order $M^2$, we obtain

\be\label{2.8} {1\over 2a}\left({\partial S_0\over\partial\phi}\right)^2 + {ka\over 2} = 0,\ee
which essentially is the Hamilton-Jacobi equation for source-free gravity. This equation reduces to the vacuum Einstein's equation $\dot a^2 + k = 0$, if we identify $\rho_a$ with $M{\partial S_0\over \partial a}$, or equivalently under the choice of the time parameter \cite{5}

\be\label{2.9} {\partial \over \partial t}=-{1\over a}{\partial S_0\over \partial a}{\partial \over \partial a}.\ee
At this juncture we recall that the validity of WKB approximation under consideration requires $|{d\lambda_a\over da}|\ll 1$, where $\lambda_a$ is the de Broglie wavelength for pure gravity. Identifying the derivative of the phase $S_0$ with the canonical momentum, we observe that the above statement reduces to $|{d\over da}\big({\hbar \over MS_0}\big)|\ll 1$. In view of equation \eqref{2.8} this implies that $a\gg\sqrt{55} \times 10^{-29}$ cm. This gives a lower limit to $a$, beyond which semiclassical approximation fails. However, it is only a few order of magnitude larger than the Planck's length and as such both microscopic and macroscopic wormholes would possibly exist. Hence this method of approximation is well posed to address both the problems of vanishing of the cosmological constant and the final stage of evaporation and complete disappearance of cosmological black holes. Although in the present work we have not studied the evolution of this finite resolution limit of the scale factor, we think that there might exist appropriate conditions under which a microscopic wormhole would evolve to a macroscopic one. Next, to the following order $M$, we obtain

\be \begin{split}\label{2.10}& {i\hbar\over 2a}{\partial^2S_0\over \partial a^2}-{1\over a}{\partial S_0\over \partial a}{\partial S_1\over\partial a}+ {i\hbar\over 2a^2}(n - 12\epsilon l){\partial S_0\over \partial a}+{6\epsilon \phi\over a^2}{\partial S_0\over \partial a}{\partial S_1\over \partial \phi}-{i\hbar\over 2a^3}{\partial^2 S_1\over \partial \phi^2}\\
& +{1\over 2a^3}\left({\partial S_1\over \partial\phi}\right)^2-{i\hbar l\over 2a^3 \phi}{\partial S_1\over \partial \phi}+6\epsilon\phi^2 (1-3\epsilon)k a + a^3 V(\phi)= 0.\end{split}\ee
Now if we define a functional $f(a,\phi)$ as $f(a,\phi) =  D(a)\exp[{iS_1(a,\phi)\over \hbar}]$, where $D(a)$ is chosen to satisfy the equation

\be \label{2.11} \left[{1\over a}{\partial S_0\over \partial a}{\partial\over \partial a}-{1\over 2}\left({1\over a}{\partial^2 S_0\over \partial a^2}+{n-12\epsilon l\over a^2}{\partial S_0\over \partial a}\right)\right] D(a) = 0,\ee
then equation \eqref{2.10} may be recast in the following form

\be \begin{split}\label{2.12}-{i\hbar\over a}\left({\partial S_0\over\partial a}\right){\partial f(a,\phi)\over \partial a}=\left[-{\hbar^2\over 2a^3}\left({\partial^2 \over \partial \phi^2}+{ l\over \phi}{\partial \over \partial\phi}\right) \pm {6\sqrt{k}\hbar\epsilon \phi\over a}\left({\partial \over \partial\phi}\right) + a^3V(\phi) +6k(1-3\epsilon)\epsilon a \phi^2\right] f(a,\phi).\end{split}\ee
Now under the choice \eqref{2.9} of the time parameter, the left hand side of equation \eqref{2.12} reduces to ${\partial f\over \partial t}$. Further, in view of the Hamiltonian \eqref{2.5} it is apparent that for minimal coupling there is no operator ordering ambiguity between $\hat \phi$ and $\hat p_{\phi}$, and so one can set $l = 0$. Hence the right hand side of \eqref{2.12} reduces to the Hamiltonian operator for the minimally coupled field operating on $f(a,\phi)$, in the background of classical curved space. Thus for $\epsilon = 0$, \eqref{2.12} reduces to the functional Schr\"odinger equation propagating in the background of classical curved space. It is now left to show that for arbitrary coupling $\epsilon \ne 0$, the right hand side of \eqref{2.12} is essentially the Hamiltonian operator for non-minimally coupled field $\phi$ operating on the function $f(a,\phi)$, in the background of classical curved space. For this pupose, we first recall that, to find the Hamiltonian for minimally coupled matter field in the background of curved space, it is required to start with the action $\int d^4x\sqrt{-g} L_m = \int\big[{1\over 2}\dot\phi^2 - V(\phi)\big]a^3 dt$, and to carry out the variation with respect to field $\phi$ keeping $a$ fixed, since $a$ has already been determined by the vacuum Einstein's equation,- $\dot a^2 + k = 0$. Likewise, here we start with the action

\be \label{2.13} A = \int d^4x \sqrt{-g}\left[{1\over 4\pi ^2}\epsilon\phi^2 R - {1\over 2\pi^2}\left({1\over 2}\phi_{,\mu}\phi_{,\nu}g^{\mu\nu}+V(\phi)\right)\right]+{1\over 2\pi^2}\int_{\Sigma}d^3 x\sqrt{h} K\epsilon\phi^2.\ee
Remember that pure gravity has already been separated following equation \eqref{2.8}, so here we have considered only the part of the action \eqref{2.1} which does not include pure gravity. In the process, action \eqref{2.13} reduces to the action for induced theory of gravity. In the Robertson-Walker line element \eqref{2.2} under consideration, and using equation \eqref{2.8}, equation \eqref{2.13} takes the form

\be\label{2.14} A \int\left[-6\epsilon a\phi\dot\phi\left({\partial S_0\over \partial a}\right) +{1\over 2}a^3\dot\phi^2 - 6\epsilon k a\phi^2 - a^3 V(\phi)\right].\ee
Hence the Hamiltonian $H = p_{\phi}\dot \phi - L$, reads

\be \label{2.15} H = {p_{\phi}^2\over 2a^3} +{6\epsilon\phi\over a^2}\left({\partial S_0\over \partial a}\right)p_{\phi} + 6k\epsilon (1-3\epsilon)a \phi^2 + a^3 V(\phi).\ee
To take up canonical quantization scheme, we must remove some of the operator ordering ambiguities between $\hat \phi$ and $\hat p_{\phi}$, by replacing $\hat p_{\phi}^2$ by $\phi^{-l}\hat p_{\phi} \phi^l \hat p_{\phi}$. The functional Schr\"odinger equation for the non-minimally coupled matter field propagating in the background of curved space then takes the form

\be \label{2.16} i\hbar {\partial f\over \partial t}=\left[-{\hbar^2\over 2a^3}\left({\partial^2\over \partial\phi^2} + {l\over \phi}{\partial\over \partial\phi}\right) \pm {6\hbar \sqrt{k}\epsilon\phi\over a}{\partial\over \partial\phi} + 6k\epsilon(1-3\epsilon) a\phi^2 +a^3 V(\phi)\right]f,\ee
which is essentially equation \eqref{2.12}. Therefore we have achieved our goal. It is now possible to integrate equation \eqref{2.11} to find the solution of $D(a)$ as

\be\label{2.17} D(a) = m a^{{1\over 2(1+n-12\epsilon l)}},\ee
$m$ being the constant of integration. Hence up-to this order of approximation $M$, the semiclassical wave-function $\psi = \exp{i\over \hbar}[MS_0+S_1]$ turns out to be

\be \label{2.18} \psi = {\exp{\left(-{M\sqrt{k}\over 2\hbar}a^2\right)}\over m a^{{1\over 2(1+n-12\epsilon l)}}}f(a,\phi).\ee
We have considered negative sign in the exponent in order to ensure that $\psi$ should exponentially damp for large three-geometry. However, the fate of the wave-function at small three-geometry remains obscure, until the solution for $f(a,\phi)$ is found in view of equation \eqref{2.12} or \eqref{2.16} as well. For this purpose, we use equation \eqref{2.8} in equation \eqref{2.12}. Thus for the closed model ($k = +1$), under the choice $f(a,\phi) =  f(\alpha)$, where, $\alpha =  a\phi$, equation \eqref{2.12} reduces, for a quartic potential $V(\phi) =  \lambda\phi^4$ to

\be\label{2.19} f_{,aa} + \left[{1\over \alpha} + {2\over \hbar}(1-6\epsilon)\alpha\right]f_{,a} - {2\over \hbar^2}[6\epsilon(1-3\epsilon)+\lambda a^2]\alpha^2 f,\ee
where $f_{,\alpha}$ stands for the derivative of $f$ with respect to $\alpha$. For the sake of convenience, only the positive sign has been considered. It should be noted that $\alpha = 0$ only gives a non-essential regular singularity. Next, we make a transformation,

\be \label{2.20} f = \alpha^{-{1\over 2}} g(\alpha) \exp{\left[-{(1-6\epsilon)\alpha^2\over 2\hbar}\right]},\ee
so that the first order derivative term in equation \eqref{2.19} does not appear. Thus we obtain

\be \label{2.21} g_{,\alpha\alpha} +\left[{l(2-l)\over 4}\alpha^{-2} -{(1-6\epsilon)(1+l)\over \hbar} -{\alpha^2\over \hbar^2} -{2\lambda\over \hbar^2}\alpha^4\right]g(\alpha) = 0.\ee
Equation \eqref{2.21} admits a series solution \[g(\alpha) = \sum_{n=0}^\infty g_n \alpha^{n+s},\]
corresponding to which we obtain a recurrence relation involving four terms as

\be\label{2.22} g_{j+6}\left[(j+s+6)(j+s+5) + {l(2-l)\over 4}\right] -{(1-6\epsilon)(1+l)\over \hbar}g_{j+4} -{1\over \hbar^2}g_{j+2} -{2\lambda\over \hbar^2}g_j = 0.\ee
Hence two sets of coefficients may be obtained in terms of $g_0$, which remains arbitrary: one set is for $s = {l\over 2}$, and the other for $s = 1-{l\over 2}$. Note that for $l = 1$, only one independent solution of equation \eqref{2.21} would emerge. The coefficients are,

\[ g_2 = {1-6\epsilon\over 2\hbar}g_0;\hspace{1.15 in}g_4 = {(1+l)(1-6\epsilon)^2 +2 \over 8\hbar^2(l+3)}g_0;\]
\be \label{2.23a} g_6 = {1\over 6(l+5)}\left[{(1+l)^2(1-6\epsilon)^3+2(1-6\epsilon)(3l+7) \over 8\hbar^3(l+3)}+{2\lambda\over \hbar^2}\right]g_0,\ee
while, $g_1 = g_3 = g_5 = ......= 0$, for $s={l\over 2}$, where $l\ne -3, -5, -7 ...$ etc. and

\[ g_2 = {1-6\epsilon \over 2\hbar}\left({1+l\over 3-l}\right) g_0;\hspace{0.4 in}g_4 = \left[{(1+l)^2(1-6\epsilon)^2+ 2(3-l)\over 8\hbar^2(5-l)(3-l)}\right]g_0;\]
\be \label{2.23b} g_6 = {1\over 6(7-l)}\left[{(1+l)^3(1-6\epsilon)^3 +2 (1+l)(13-3l)(1-6\epsilon)\over 8\hbar^3(3-l)(5-l)}+{2\lambda\over\hbar^2}\right]g_0,\ee
etc. while $g_1 = g_3 = g_5 = ......= 0$, for $s=1-{l\over 2}$, where $l\ne 3, 5, 7 ...$ etc. Now for the set of coefficients \eqref{2.23a}, the wave-function $\psi$ of \eqref{2.18} takes the form in view of the transformation \eqref{2.20}

\be \begin{split}\label{2.24}& \psi = {g_0\over m}\left[{\exp{\left(-{M a^2\over 2\hbar}\right)}\exp{\left(-{(1-6\epsilon)a^2\phi^2 \over 2\hbar}\right)}\over a^{{(1+n-12\epsilon l)\over 2}}}\right]\times\Bigg[1+{1-6\epsilon\over 2\hbar}a^2\phi^2+{(1+l)(1-6\epsilon)^2+2\over 8\hbar^3(l+3)}\\
&+{1\over 6(l+5)}\times\left\{{(1+l)^2(1-6\epsilon)^3 + 2(3l+7)(1-6\epsilon)\over 8\hbar^3(l+3)}+{2\lambda\over \hbar^2}\right\}a^6\phi^6 + .......\Bigg]
\end{split}\ee
The above wave-function will be damped at $a\rightarrow \infty$, if $\epsilon$ is not too large a positive quantity. If $\epsilon\gg {1\over 6}$, then the argument of the second exponent becomes positive. The series in that case might play the dominant role, for which the wave-function might diverge for large value of $a$. As $a\rightarrow 0$, the numerator becomes a constant, and so the denominator decides the behaviour of $\psi$. For $l = 0$, a wormhole solution is admissible for $n = -1$, irrespective of the sign of $\epsilon$. Hence, we find that even for large negative value of $\epsilon (\sim 10^{-3})$, as required for obtaining the observed order of magnitude of density perturbation ${\partial\rho\over \rho}\sim 10^{-4}$, keeping $\lambda$ within the GUT range $(\sim 10^{-2})$, a wormhole configuration exists. This is definitely a new wormhole solution. In the conformally coupled case, $\epsilon = {1\over 6}$, the Hawking-Page \cite{25} wormhole solution is recovered for $l=1$ and $n=1$. For large negative $\epsilon$, the wave-function $\psi$ diverges as $a\rightarrow 0$, for $l >0$. One can in that case choose $l < 0$, to regulate $\psi$. However, if one considers the validity range of semiclassical approximation ($a \gg \sqrt{55}\times 10^{-29}$ cm.), then at the limit, one might obtain other wormhole configurations, as $\psi$ would remain constant in the limit, below which $\psi \rightarrow \infty$. This was speculated by Coule \cite{20}. It is to be noted that the situation remains unaltered even for $\lambda = 0$, i.e. for free field.\\

In a similar manner it is possible to construct the wave-function for the set of coefficients \eqref{2.23b}. In that case,

\be \begin{split}\label{2.25}& \psi = {g_0\over m}\left[{\exp{\left(-{M a^2\over 2\hbar}\right)}\exp{\left(-{(1-6\epsilon)a^2\phi^2 \over 2\hbar}\right)\phi^{(1-l)}}\over a^{{(n-1+2l(1-6\epsilon)\over 2}}}\right]\\&\times\Bigg[1+{1-6\epsilon\over 2\hbar}\left({1+l\over 3-l}\right)a^2\phi^2+{(1+l)^2(1-6\epsilon)^2+2(3-l)\over 8\hbar^2(5-l)(3-l)}a^4\phi^4\\
&+{1\over 6(7-l)}\left\{{(1+l)^3(1-6\epsilon)^3 + 2(1+l)(13-3l)(1-6\epsilon)\over 8\hbar^3(3-l)(5-l)}+{2\lambda\over \hbar^2}\right\}a^6\phi^6 + .......\Bigg]
\end{split}\ee
As before, if $\psi$ is not too large a positive quantity, then $\psi \rightarrow 0$ as $a \rightarrow \infty$ and $\psi \rightarrow$ constant as $a \rightarrow 0$ for $l = 0$ and $n = 1$. This is yet another wormhole configuration that exists even for arbitrary large negative value of $\epsilon$. For conformal coupling $\epsilon = {1\over 6}$, the Hawking-Page wormhole configuration \cite{25} can be recovered for $n = 1$, even without restricting $l$. Wormhole configuration may be obtained as before for large negative value of $\epsilon$ and $l < 0$, if one restricts the scale factor to $a \gg \sqrt{55}\times 10^{-29}$ cm., which is the limit for the validity of semi-classical approximation, as already mentioned. The situation here too remains unaltered for free scalar field $\lambda = 0$.

\section{Conclusion}

The WKB phase for the Wheeler-DeWitt wave-functional for gravity with non-minimally coupled scalar field has been expanded in the inverse power series of the Planck's mass to obtain the semi-classical wave-functional. In the process, the Tomonaga-Schwinger equation has been obtained and solved for quartic potential. The wave-functional has been found to produce the Hawking-Page wormhole solution for the conformally coupled scalar field $\epsilon = {1\over 6}$. Special attention has been paid in studying the case of arbitrarily large negative $\epsilon$. This is because, the recent results of Fakir, Unruh and Habib have explored the fact that, large negative $\epsilon$ can give rise to well-behaved, self-consistent classical solutions, which admit inflationary behaviour, and within which in a certain class of cases, the density perturbation problem may be solved. Our analysis has revealed that at least one wormhole solution exists even for arbitrarily large negative value of $\epsilon$. The possibility of obtaining a class of wormhole solutions for large negative $\epsilon$ has also been explored. Further, it has been noted that for every large positive $\epsilon$, the wave-functional might possibly diverge, as the scale factor goes over to infinity. This definitely gives a further physical motivation for considering large negative values of the non-minimal coupling constant. It is noteworthy that the forms of the wave-functional \eqref{2.24} and \eqref{2.25} remain unaltered for $\lambda = 0$. Hence the wormhole configurations which have been found to exist in the case of quartic potential will also exist even for a free field.\\

\noindent
\textbf{Acknowledgement}\\
Thanks are due to the referee for his useful comments.

\end{document}